\documentstyle[12pt,a4]{article}

\def\nm{\nonumber}
\def\pa{\partial}
\def\beqa{\begin{eqnarray}}
\def\beq{\begin{equation}}
\def\eeqa{\end{eqnarray}}
\def\eeq{\end{equation}}
\def\lab{\label}
\def\ovM{\overline{M}}
\def\ovN{\overline{N}}
\def\ovm{\overline{\mu}}
\def\ovn{\overline{\nu}}

\def\M{\cal{M}}

\def\O{\cal{O}}
\def\D{\cal{D}}
\def\F{\cal{F}}
\def\A{\cal{A}}
\def\K{\cal{K}}

\def\RR{\mbox{\boldmath$R$}^2}
\def\RRR{\mbox{\boldmath$R$}^3}
\def\RRRR{\mbox{\boldmath$R$}^4}

\begin{document}

\begin{titlepage}
\vspace*{-3.0cm}
\begin{center}
\begin{tabular}{c}
\makebox[15.5cm]{\large\sf Hiroshima University} \\ 
\hline
\end{tabular}
\end{center}
\vspace{1.0cm}
\begin{center}
\begin{tabular}{|c|}\hline
 \\
{\Large \bf 
Topological Field Theories }\\
 \\
{\Large \bf 
associated with } \\
 \\
{\Large \bf 
Three Dimensional} \\
\\
{\Large \bf
Seiberg-Witten Monopoles}\\
 \\ 
\hline
\end{tabular}
\end{center} 

\lineskip .80em
\vskip 2em
\normalsize
\begin{center}
{\large Y\H uji Ohta}
\vskip 1.0em
{Department of Mathematics \\
Faculty of Science\\
Hiroshima University \\
Higashi-Hiroshima 739, Japan.}
\vskip 1.0em
\end{center}
\vskip2em
\begin{center}
\begin{tabular}{c}\hline
\makebox[15.5cm]{ } \\ 
\end{tabular}
\end{center}
\vspace{-1.2cm}
\begin{abstract}
Three dimensional topological field theories associated with the three 
dimensional 
version of Abelian and non-Abelian Seiberg-Witten monopoles are presented. 
These three dimensional monopole equations are obtained by a 
dimensional reduction of the four dimensional ones. The starting actions to be 
considered are Gaussian types with random auxiliary fields. As 
the local gauge symmetries with topological shifts 
are found to be first stage reducible, Batalin-Vilkovisky algorithm 
is suitable 
for quantization. Then BRST transformation rules are automatically obtained. 
Non-trivial observables associated with 
Chern classes are obtained from geometric sector and are found to 
correspond to those of the topological field theory of Bogomol'nyi monopoles. 

\end{abstract}
\begin{center}
\begin{tabular}{c}\hline
\makebox[15.5cm]{ } \\ 
\end{tabular}
\end{center}

\end{titlepage}

\pagestyle{myheadings}
\markboth{Topological field theories associated with 3-d SW monopoles} 
{Topological field theories associated with 3-d SW monopoles}

\baselineskip 0.8cm

\begin{center}
\section{Introduction}
\end{center}

\renewcommand{\theequation}{1.\arabic{equation}}\setcounter{equation}{0}

Topological field theories \cite{BBRT,T2,S,W1} are often used to study 
topological nature of manifolds. In particular, three and four dimensional 
topological field theories are well developed. 
The most well-known three dimensional topological field theory would be 
the Chern-Simons theory whose partion function gives Ray-Singer 
torsion of three manifolds \cite{S} 
and the other topological invariants can be obtained as gauge invariant 
observables i.e., Wilson loops. The correlation functions can be 
identified with knot or link invariants e.g., Jones polynomal or its 
generalizations. On the other hand, in four dimensions, a twisted $N=2$ 
supersymmetric Yang-Mills theory developed by Witten \cite{W1} 
also has a nature of topological field theory. This Yang-Mills theory 
can be interpreted as Donaldson theory \cite{DD} and 
the correlation functions are identified with Donaldson polynomials 
which classify smooth structures of topological four manifolds. 
However, a new topological field theory on four manifolds was discovered 
in recent studies of electric-magnetic duality of supersymmetric 
gauge theory. The story of this is described as follows. 

Seiberg and Witten \cite{SW1,SW2} studied the electric-magnetic duality 
of $N=2$ supersymmetric $SU(2)$ Yang-Mills gauge 
theory (for reviews, see Refs. 8,9,10,11,12) by using a version of 
Montonen-Olive duality and obtained exact solutions. 
According to this result, the exact 
low energy effective action can be determined by a certain elliptic curve 
with a parameter $u=\langle \mbox{tr}\phi^2 \rangle$, where $\phi$ is a 
complex scalar field in the adjoint representation of the gauge group, 
describing the quantum moduli space. For large $u$, the theory is weakly 
coupled and semiclassical, but at $u=\pm \Lambda^2$ corresponding to 
strong coupling regime, where $\Lambda$ is the 
dynamically generated mass scale, the elliptic curve becomes singular 
and the situation of the theory changes drastically. At these singular 
points, magnetically charged particles become massless. Witten 
showed that at $u=\pm\Lambda^2$ 
the topological quantum field theory was related to the moduli problem 
of counting the solution of the (Abelian) ``Seiberg-Witten monopole 
equations'' \cite{W2} and it gave a dual description for 
the $SU(2)$ Donaldson theory. The particulary interesting fact is that 
the partition function of this $U(1)$ gauge 
theory produces a new topological invariant \cite{W2,W4,A,D,Brad,KM,Tau}.

The topological field theory of the Seiberg-Witten 
monopoles was discussed by several authors. 
Labastida and Mari\~{n}o \cite{LM1} took the Mathai-Quillen 
formalism \cite{MQ,AJ,Blau} and found that the resulting action 
was equivalent to that of the twisted $N=2$ supersymmetric Maxwell 
coupled with a twisted $N=2$ hypermultiplet. Furthermore, they 
generalized their results for non-Abelian cases \cite{LM2,LM3} 
and determined polynomial invariants for $SU(2)$ case corresponding 
to a generalization of Refs. 13,14 in Abelian case. In these studies, 
the topological field theories were formulated as Witten type. 
On the other hand, Hyun {\em et al.} \cite{HPP1,HPP2} discussed 
a non-Abelian topological field theory in view of twisting of $N=2$ 
supersymmetric Yang-Mills coupled with $N=2$ matters and obtained similar 
polynomial invariants. 
There are other approaches to obtain the topological action, in fact, 
Zang {\em et al.} \cite{ZWCM} derived the topological action 
as a BRST variation of a certain gauge fermion 
and Gianvittorio {\em et al.} \cite{GMR1,GMR2} discussed in view of 
a covariant gauge fixing procedure.

In three dimensions, a topological field theory of Bogomol'nyi monopoles 
can be obtained from a dimensional reduction of Donaldson 
theory \cite{BRT,BG} and 
the partition function of this theory gives the Casson invariant \cite{AJ}. 
However, the three or two dimensional topological field theory of 
Seiberg-Witten monopoles does not seem to be fully discussed, although several 
authors point out its importance \cite{T1,ZWCM,O}. 
Zang {\em et al.} \cite{ZWCM} 
performed a dimensional reduction of the Abelian Seiberg-Witten theory 
from four to three dimensions and found the reduced topological action. 
They also found in view of Mathai-Quillen formalism 
that the partition function of this three dimensional theory can 
be interpreted as a Seiberg-Witten version of the 
Casson invariant of three manifolds. 

In this paper, we discuss the topological quantum field theories associated 
with the three dimensional version of Abelian and non-Abelian Seiberg-Witten 
monopoles 
by applying Batalin-Vilkovisky quantization. 
In particular, we construct the topological actions, topological observables 
and BRST transformation rules. In section 2, we briefly review the essence 
of topological 
quantum field theories both Witten type and Schwarz type. 
The reader who is interested in the results of this paper may neglect this section. 
In section 3, 
the dimensional reduction of the Abelian and non-Abelian 
Seiberg-Witten monopole equations are explicitly performed 
and the three dimensional monopole (3-d monopole) equations are obtained. 
We also obtain quadratic actions which reproduce these three dimensional 
monopole equations as minimum. In section 4, we construct 
topological field theories of these three 
dimensional monopoles taking the actions including random 
auxiliary fields as a starting point. As the local gauge symmetries of 
them are classified 
as first stage reducible with on-shell reducibility, Batalin-Vilkovisky 
algorithm is suitable to quantize these theories. 
Then we can automatically obtain the BRST transformation rules by 
construcion. It is shown 
that the observables in geometric sector can be obtained from a standard 
fashion, but those in matter sector are found to be trivial. 
The reader will find that our results for Abelian case are consistent with 
those of the dimensionally reduced version of the topological 
field theory of four dimensional Seiberg-Witten 
monopoles \cite{ZWCM}, while those 
for non-Abelian case are new results. 
It is interesting to compare our results with 
those of the topological field theory of 
Bogomol'nyi monopoles. Section 5 is a summary 
and we mention some open problems. In Appendix A and B, we summarize the 
result of 
the Batalin-Vilkovisky quantization for the four dimensional non-Abelian 
Seiberg-Witten monopoles, for the reader's convenience. 

\begin{center}
\subsection*{Notations}
\end{center}

We use following notations unless we mention especially. 
Let $X$ be a compact 
orientable spin four manifold with no boundary and 
$g_{\mu\nu}$ be its Riemannian metric tensor with $g=\mbox{det}\ g_{\mu\nu}$. 
We use $x_{\mu}$ as the local coordinates on $X$. 
$\gamma_{\mu}$ are Dirac's gamma matrices and $\sigma_{\mu\nu}=[\gamma_{\mu},
\gamma_{\nu}]/2$ with $\{\gamma_{\mu},\gamma_{\nu}\}=g_{\mu\nu}$ (see also 
Appendix C). 
$M$ is a Weyl fermion and $\ovM$ is a complex conjugate 
of $M$. We suppress spinor indices. The Lie algebra {\bf g} is 
defined by $[T^a , T^b ]=if_{abc} T^c $, 
where $T^a$ is a generator normalized as $\mbox{tr }T^a T^b =\delta^{ab}$. 
The symbol $f_{abc}$ is a structure constant of {\bf g} and is anti-symmetric 
in its indices. 

The Greek indices $\mu,\nu,\alpha$ etc run from 
0 to 3. The Roman indices 
$a,b,c,\cdots$ are used for the Lie algebra indices running from 1 to 
dim {\bf g}, whereas $i,j,k,\cdots$ 
are the indices for space coordinates. Space-time indices are raised and 
lowered with $g_{\mu\nu}$. The repeated indices are assumed to be summed. 
$\epsilon_{\mu\nu\rho\sigma}$ is an anti-symmetric tensor with 
$\epsilon_{0123}=1$. We often use the abbreviation of roman indices as  
$\theta =\theta^a T^a$ etc in order to suppress the summation over 
Lie algebra indices.

\begin{center}
\section{Quick tour to topological field theory}
\end{center}

\renewcommand{\theequation}{2.\arabic{equation}}\setcounter{equation}{0}

This section is devoted to a brief review of topological field theory. 
The reader who is interested in the details should refer 
to Refs. 1,2,3,4,9. 

Let $\phi$ be any field content. For a local symmetry of $\phi$, we can 
construct a nilpotent BRST operator $Q_B$ $(Q_{B}^2 =0)$. The 
variation of any functional ${\O}$ of $\phi$ is denoted by 
	\beq
	\delta {\O}=\{ Q_B ,{\O}\},
	\eeq
where the blacket $\{*,*\}$ means a graded commutator, namely, if ${\O}$ is 
bosonic the bracket means a commutator $[*,*]$ and otherwise it is 
an anti-bracket.

Then we can give the definition of topological field theory \cite{BBRT}. 

{\bf Definition} {\em A topological field theory consists of 
\begin{enumerate}
\item
a collection of Grassmann graded fields $\phi$ on an $n$-dimensional 
Riemannian manifold $X$ with a metric $g$,
\item
a nilpotent Grassmann odd operator $Q$,
\item
physical states to be $Q$-cohomology classes,
\item
an energy-momentum tensor $T_{\alpha\beta}$ which is $Q$-exact for 
some functional $V_{\alpha\beta}$ such as 
	\beq
	T_{\alpha\beta}=\{Q,V_{\alpha\beta}(\phi,g)\}.
	\eeq
\end{enumerate}}

In this definition, $Q$ is often identified with $Q_B$ and is 
in general independent of the metric. There are several examples 
of topological field theories which do not satisfy this 
definition, but this definition is useful in many cases. 

There are two broad types of topological field theories satisfying 
this definition and they are classified into Witten type \cite{W1} 
or Schwarz type \cite{S} (there are several non-standard Schwarz type 
theories, 
e.g., higher dimensional BF theories, but here we do not consider such cases). 

For Witten type theory, the quantum action $S_q$ which comprises the 
classical action, ghost and gauge fixing terms, can be represented by 
$S_q =\{Q_B ,V\}$, for some function $V$ of metric and fields and 
BRST charge $Q_B$. 
Under the metric variation $\delta_g$ of the partition function $Z$, 
it is easy to see that 
	\beqa
	\delta_g Z &=&\int {\D}\phi \ e^{-S_q}\left(
	-\frac{1}{2}\int_{X}d^n x \sqrt{g}\delta g^{\alpha\beta}
	T_{\alpha\beta}\right)\nm\\
	&=&\int {\D}\phi \ e^{-S_q}\{ Q,\chi \}\nm\\
	&\equiv&\langle \{ Q,\chi\} \rangle =0,
	\lab{va}
	\eeqa
where 
	\beq
	\chi =-\frac{1}{2}\int_{X}d^n x \sqrt{g}\delta 
	g^{\alpha\beta}V_{\alpha\beta}
	.\eeq
The last equality in (\ref{va}) follows from the BRST invariance of the 
vacuum and means that $Z$ is independent of the local structure of $X$, 
that is, $Z$ is a ``topological invariant'' of $X$.  

In general, for Witten type theory, $Q_B$ can be constructed by 
an introduction of a topological 
shift with other local gauge symmetry \cite{BS,BMS}. For example, in order to 
obtain the topological Yang-Mills theory on four manifold $M^4$, we 
introduce the shift in the gauge transformation 
for the gauge field $A_{\mu}^a$ such as $\delta A_{\mu}^a =D_{\mu}\theta^a +
\epsilon_{\mu}^a$, where $D_{\mu}$ is a covariant derivative, $\theta^a$ 
and $\epsilon_{\mu}^a$ are the (Lie algebra valued) usual gauge transformation 
parameter and topological shift parameter, respectively. In order to see the 
role of this shift, let us consider the first Pontrjagin class 
on $M^4$ given by
	\beq
	S=\frac{1}{8}\int_{M^4}\epsilon^{\mu\nu\rho\sigma}F_{\mu\nu}^a 
	F_{\rho\sigma}^a d^4 x 
	\lab{clasac}
	,\eeq
where $F_{\mu\nu}^a$ is a field strength of the gauge field. 
We can easily check the invariance of (\ref{clasac}) under the action of 
$\delta$. In this sense, (\ref{clasac}) has a larger 
symmetry than the usual (Yang-Mills) gauge symmetry. 
Taking this into account, we can construct the topological Yang-Mills gauge 
theory \cite{BS,BMS,GPS}. We can also consider similar ``topological'' 
shifts for matter fields as will be shown in section 4. 

In addition, in general, Witten type topological field theory can be 
obtained from the quantization of some Langevin equations \cite{BRT}. 
This approach has been used for the construction of several topological field 
theories, e.g., supersymmetric quantum mechanics, topological sigma models or 
Donaldson theory \cite{BBRT,LP} (we will use this approach for the 
$N=4$ theory \cite{VW} in next communication \cite{Ohta}).

On the other hand, Schwarz type theory \cite{S} begins with any metric 
independent classical action $S_{c}$ as a starting point, but $S_{c}$ is 
assumed not to be a total derivative. Then the quantum action (up to gauge 
fixing term) can be written by 
	\beq
	S_{q}=S_{c} +\{Q,V(\phi ,g)\}
	\lab{shw}
	\eeq
for some function $V$. For this quantum action, we can easily check the 
topological nature of the partition function, but note that the 
energy-momentum tensor contributes only from the second term in (\ref{shw}). 
One of the differences between Witten type and 
Schwarz type theories can be seen in this point. Namely, the energy-momentum 
tensor of the classical action for Schwarz type theory vanishes because 
it is derived as a result of metric variation. 

Finally, we comment on the local symmetry of Schwarz type theory. Let us 
consider the Chern-Simons theory as an example. The classical action 
	\beq
	S_{CS}=\int_{M^3}d^3 x \left( A\wedge dA+\frac{2}{3}A\wedge A\wedge 
	A\right)
	\eeq
is a topological invariant which gives the second Chern class of three 
manifold $M^3$. As is easy to find, $S_{CS}$ is not invariant 
under the topological gauge transformation, although it is (Yang-Mills) 
gauge invariant. Therefore the quantization is proceeded by the 
standard BRST method. This is a general feature of Schwarz type theory.

\begin{center}
\section{Dimensional reduction}
\end{center}

\renewcommand{\theequation}{3.\arabic{equation}}\setcounter{equation}{0}

In this section, the dimensional reduction of the Abelian and 
non-Abelian Seiberg-Witten monopole 
equations is presented. For mathematical progresses on Seiberg-Witten 
monopoles, see Refs. 15,16 for Abelian case 
and Ref. 17 for non-Abelian case.

First, let us recall the Seiberg-Witten monopole equations in four dimensions. 
We assume that $X$ has Spin structure. Then there exist 
rank two positive and negative spinor bundles $S^{\pm}$. For 
Abelian gauge theory, we introduce a complex line bundle 
$L$ and a connection $A_{\mu}$ on $L$. The Weyl spinor $M(\ovM)$ is 
a section of $S^+ \otimes L$ $(S^+ \otimes L^{-1})$, hence $M$ satisfies 
the positive chirality condition $\gamma^5 M=M$. If $X$ does not 
have Spin structure, we introduce $\mbox{Spin}^{\mbox{\scriptsize \bf c}}$ 
structure and $\mbox{Spin}^{\mbox{\scriptsize \bf c}}$ bundles 
$S^{\pm} \otimes L$, where $L^2$ is a line bundle. 
In this case, $M$ should be interpreted as a section of 
$S^+ \otimes L$. Below, we assume Spin structure. 
The reader who is interested in the physical implications of Spin 
and $\mbox{Spin}^{\mbox{\scriptsize \bf c}}$ 
structures should refer to the excellent review Ref. 11 and references 
therein.

The Abelian Seiberg-Witten monopole equations \cite{W2} in four dimensions 
are the set of following differential equations
	\beqa
	F_{\mu \nu}^{+} +\frac{i}{2}\ovM \sigma_{\mu \nu}M &=&0 ,\nm \\
	i\gamma^{\mu}D_{\mu}M &=&0 ,
	\lab{mono1}
	\eeqa
where $F_{\mu \nu}^{+}$ is the self-dual part of the $U(1)$ curvature tensor  
	\beqa
	F_{\mu\nu} &=&\pa_{\mu} A_{\nu} -\pa_{\nu}A_{\mu},\nm \\ 
	F_{\mu \nu}^{+} &=& P_{\mu\nu\rho\sigma}^{+}F^{\rho\sigma}
	\lab{tens}
	\eeqa
and $P_{\mu\nu\rho\sigma}^{+}$ is the self-dual projector defined by 
	\beq
	P_{\mu\nu\rho\sigma}^{+}=\frac{1}{2}\left( \delta_{\mu\rho}
	\delta_{\nu\sigma} 
	+\frac{\sqrt{g}}
	{2}\epsilon_{\mu \nu \rho \sigma}\right).	
	\eeq
Note that the second term in the first equation of (\ref{mono1}) 
is also self-dual \cite{T1}. On 
the other hand, the second equation in (\ref{mono1}) 
is a twisted Dirac equation whose covariant derivative $D_{\mu}$ is given by 
	\beq
	D_{\mu} =\pa_{\mu} +\omega_{\mu} -iA_{\mu}
	,\eeq
where
	\beq
	\omega_{\mu}=\frac{1}{4}\omega_{\mu}^{\alpha \beta}[
	\gamma_{\alpha} ,\gamma_{\beta} ]
	\eeq
is the spin connection 1-form on $X$. 

In order to perform a reduction to three dimensions, let us first assume that 
$X$ is a product manifold of the form $X=Y\times [0,1]$, 
where $Y$ is a three dimensional compact manifold which 
has Spin structure. We may identify $t \in [0,1]$ as 
a ``time'' variable, or, we assume $t$ as the zero-th coordinate of 
$X$, whereas $x_i$ $(i=1,2,3)$ are the coordinates on (space manifoild) $Y$. 
Then the metric is given by 
	\beq
	ds^2 =dt^2 +g_{ij}dx^i dx^j 
	.\eeq
The dimensional reduction is proceeded by assumnig that all fields are 
independent of $t$. Below, we suppress the volume factor $\sqrt{g}$ of $Y$ 
for simplicity. 

First, let us consider the Dirac equation. After the dimensional reduction, 
the Dirac equation will be 
	\beq
	\gamma^i D_i M-i\gamma^0 A_0 M =0
	.\eeq

As for the first monopole equation, using (\ref{tens}) we find that 
	\beqa
	F_{i0}+\frac{1}{2}\epsilon_{i0jk}F^{jk}&=&-i\ovM \sigma_{i0}M,\nm\\
	F_{ij}+\epsilon_{ijk0}F^{k0}&=&-i\ovM \sigma_{ij}M
	.\lab{redself1}
	\eeqa
Since the above two equations are dual each other, the first one, 
for instance, can be reduced to the second one by a contraction 
with the totally anti-symmetric tensor. Thus it is sufficient to 
consider one of them. Here, we take the first equation in (\ref{redself1}). 

After the dimensional reduction, (\ref{redself1}) will be 
	\beq
	\pa_i A_0 -\frac{1}{2}\epsilon_{ijk}F^{jk}=-i\ovM \sigma_{i0} M
	,\lab{meq}
	\eeq 
where we have set $\epsilon_{ijk}\equiv\epsilon_{0ijk}$. 

Therefore, the three dimensional version of the Seiberg-Witten 
equations are given by  
	\beqa
	\pa_i b-\frac{1}{2}\epsilon_{ijk}F^{jk}+i\ovM\sigma_{i0} M&=&0,\nm\\
	i(\gamma^i D_i -i\gamma^0 b)M&=&0
	,\lab{orig}
	\eeqa
where $b \equiv A_0$. The factor $i$ of the Dirac equation is for later 
convenience. 

It is now easy to establish the non-Abelian 3-d monopole equations 
(for the four dimensional version, see Refs. 9,24,25,26,27 and 
Appendix A) as
	\beqa
	\pa_i b^a +f_{abc}A_{i}^b b^c -\frac{1}{2}\epsilon_{ijk}F^{ajk} +i\ovM 
	\sigma_{i0} T^a M &=&0,\nm\\
	i(\gamma^i D_i -i\gamma^0 b)M&=&0
	,\eeqa
where we have abbreaviated $\ovM \sigma_{\mu\nu}T^a M \equiv \ovM^i 
\sigma_{\mu\nu}(T^a )_{ij}M^j$, subscripts of $(T^a )_{ij}$ 
run 1 to dim {\bf g} and $b^a \equiv A_{0}^a$. 

Next, let us find an action which produces (\ref{orig}). 
We can easily find that the simplest one is given by 
	\beq
	S=\frac{1}{2}\int_{Y} 
	\left[\left(\pa_i b-\frac{1}{2}\epsilon_{ijk}F^{jk}+i \ovM 
	\sigma_{i0}M\right)^2 +
	|i(\gamma^i D_i -i\gamma^0 b)M|^2 \right]d^3 x 
	.\lab{qua}
	\eeq 
Note that the minimum of (\ref{qua}) is given by (\ref{orig}). In this sense, 
the 3-d monopole equations are not equations of motion but constraints. 
Furthermore, there is a constraint for $b$. To see this, 
let us rewrite (\ref{qua}) as 
	\beq
	S=\int_Y d^3 x \left[\frac{1}{2}\left(\frac{1}{2}\epsilon_{ijk}
	F^{jk}-i\ovM \sigma_{i0}M \right)^2 +\frac{1}{2}|\gamma^i D_i M|^2 
	+\frac{1}{4}(\pa_i b)^2 +\frac{1}{2}b^2 |M|^2 \right] 
	.\eeq
The minimum of this action is clearly given by the 3-d monopole 
equations with $b=0$, for non-trivial $A_i $ and $M$. However, 
for trivial $A_i$ and $M$, we may relax the condition $b=0$ 
to $\pa_i b=0$, i.e., $b$ is (in general) a non-zero constant. 
This can be also seen from (\ref{meq}). Accordingly, we obtain 
	\beqa
	\frac{1}{2}\epsilon_{ijk}F^{jk}-i\ovM \sigma_{i0}M&=&0,\nm\\
	i\gamma^i D_i M &=&0,\nm\\
	b =0 \ \mbox{or} \ \pa_i b&=&0,
	\lab{threemono}
	\eeqa
as an equivalent expression to (\ref{orig}), but we will use (\ref{orig}) 
for convenience. The Gaussian action will be used in the 
next section in order to construct a topological field theory by 
Batalin-Vilkovisky quantization algorithm. The non-Abelian 
version of (\ref{qua}) and (\ref{threemono}) would be obvious. 

There is another action which can produce (\ref{threemono}) as 
equations of motion. It is given by a Chern-Simons 
action coupled with a matter \cite{D,KM,ZWCM} which is analogous to the 
action in massive gauge theory \cite{DJT}, but we do not discuss the 
quantum field theory of this Chern-Simons action.

\begin{center}
\section{Topological field theories of the 3-d monopoles}
\end{center}

In this section, we construct topological field theories associated with the 
Abelian and non-Abelian 3-d monopoles by Batalin-Vilkovisky 
quantization algorithm. 

\begin{center}
\subsection{Abelian case}
\end{center}

A three dimensional action for the Abelian 3-d monopoles was found by the 
direct dimensional reduction of the four dimensional 
one \cite{ZWCM,O}, but 
we show that the three dimensional topological action can be also directly 
constructed from the 3-d monopole equations.

\begin{center}
\subsubsection{Topological action}
\end{center}

\renewcommand{\theequation}{4.\arabic{equation}}\setcounter{equation}{0}

A topological Bogomol'nyi action was constructed by using 
Batalin-Vilkovisky quantization algorithm \cite{BRT} (
similar construction can be found in two dimensional version \cite{ST}) or 
quantization of a magnetic charge \cite{BG}. The former is 
based on the quantization of a certain Langevin equation (``Bogomol'nyi 
monopole equation'') and the classical 
action is quadratic, but the latter is based on the ``quantization'' of 
the pure topological invariant by using the Bogomol'nyi monopole equation as a 
gauge fixing condition. 

In order to compare the action to be constructed with those of 
Bogomol'nyi monopoles \cite{BRT,BG}, we take Batalin-Vilkovisky 
procedure. The reader who is unfamiliar with this construction may consult 
the references \cite{BBRT,BRT,GPS,LP,ST,BV1,BV2,BV3,HM}.

In order to obtain the topological action associated with 3-d monopoles, 
we introduce random Gaussian fields $G_i$ and $\nu(\ovn)$ and 
then start with the action
	\beq
	S_c =\frac{1}{2}\int_{Y} 
	\left[\left(G_i-\pa_i b+\frac{1}{2}\epsilon_{ijk}F^{jk}-i\ovM
	\sigma_{i0}M\right)^2 +
	|(\nu -i\gamma^i D_i M-\gamma^0 bM)|^2 \right]d^3 x 
	.\lab{lag}
	\eeq 
Note that $G_i$ and $\nu (\ovn )$ are also regarded as auxiliary fields. 
This action reduces to (\ref{qua}) in the gauge 
	\beq
	G_i=0,\ \nu =0.
	\lab{con}
	\eeq

Firstly, note that (\ref{lag}) is invariant under the topological gauge 
transformation
	\beqa
	\delta A_i &=& \pa_i \theta +\epsilon_i ,\nm\\
	\delta b &=&\tau ,\nm\\
	\delta M&=& i\theta M+\varphi ,\nm\\
	\delta G_i&=&\pa_i \tau-\epsilon_{ijk}\pa^j \epsilon^k +i
	(\overline{\varphi}\sigma_{i0}M+\ovM\sigma_{i0}\varphi),\nm\\
	\delta \nu&=& i\theta\nu+\gamma^i \epsilon_i M+i\gamma^i D_i \varphi +
	\gamma^0 b\varphi +\gamma^0 \tau M 
	,\lab{alg}
	\eeqa
where $\theta$ is the parameter of gauge transformation, $\epsilon_i$ 
and $\tau \equiv \epsilon_4$ 
are parameters which represent the topological shifts and $\varphi$ 
the shift on the spinor space. The brackets for indices 
means anti-symmetrization, i.e., 
	\beq
	A_{[i}B_{j]}=A_i B_j -A_j B_i 
	.\eeq

Here, let us classify the gauge algebra (\ref{alg}). This is necessary to use 
Batalin-Vilkovisky algorithm. Let us recall that the local symmetry 
for fields $\phi_i$ can be written generally in the form 
	\beq
	\delta \phi_i =R_{\alpha}^i (\phi) \epsilon^{\alpha},
	\eeq
where the indices mean the label of fields and 
$\epsilon^{\alpha}$ is a some local parameter. 
When $\delta \phi_i =0$ for non-zero $\epsilon^{\alpha}$, this 
symmetry is called first stage reducible. 
In the reducible theory, we can find zero-eigenvectors $Z_{a}^{\alpha}$ 
satisfying $R_{\alpha}^i Z_{a}^{\alpha}=0$. Moreover, when the theory is 
on-shell reducible, we can find such eigenvectors by using 
equations of motion. 

For the case at hand, under the identifications
	\beq
	\theta=\Lambda,\ \epsilon_{i}=-\pa_i \Lambda,\ \varphi=-i\Lambda M
	\eeq
and 
	\beq
	\tau =0 \lab{tau}
	,\eeq
(\ref{alg}) will be 
	\beqa
	\delta A_i &=& 0,\nm\\
	\delta b &=&0,\nm\\
	\delta M&=& 0,\nm\\
	\delta G_i&=&0,\nm\\
	\delta \nu&=& i\Lambda(\nu-i\gamma^i D_i M -\gamma^0 bM)|_{\mbox{
	\scriptsize on-shell}}=0
	.\lab{alg2}
	\eeqa
Then for $\delta A_i$, for example, the $R$ coefficients and the 
zero-eigenvectors are derived from  	
	\beq
	\delta A_i =R_{\theta}^{A_i}Z_{\Lambda}^{\theta}+
	R_{\epsilon_j}^{A_i} Z_{\Lambda}^{\epsilon_j}=0,
	\eeq
that is 
	\beq
	R_{\theta}^{A_i}=\pa_i ,\ R_{\epsilon_j}^{A_i}=\delta_{ij},\  
	Z_{\Lambda}^{\theta}=1 ,\ Z_{\Lambda}^{\epsilon_j}=-\pa_j
	.\eeq
Of course, similar relations hold for other fields. The reader may think 
that the choice (\ref{tau}) is not suitable as a first stage reducible 
theory, but note that the zero-eigenvectors appear on every 
point where the gauge equivalence and the topological shift happen to 
coincide. 
In this three dimensional theory, $b (A_0 )$ is invariant for the usual 
infinitesimal gauge transformation because of its ``time'' independence, 
so (\ref{tau}) means that the existence of the points on spinor space 
where the topological shift trivializes indicates the first stage 
reducibility. 

If we carry out BRST quantization via Faddeev-Popov 
procedure in this situation, the Faddeev-Popov determinant will have zero 
modes. Therefore in order to fix the gauge further we need a ghost for ghost. 
This reflects on the second generation gauge invariance (\ref{alg2}) realized 
on-shell. However, since $b$ is irrelevant to $\Lambda$, the ghost for $\tau$ 
will not couple to the second generation ghost. With this in mind, 
we use Batalin-Vilkovisky algorithm in order to make BRST 
quantization. 

Let us assign new ghosts carrying opposite statistics to the local 
parameters. The assortment is given by 	
	\beq
	\theta \longrightarrow c,\ \epsilon_i \longrightarrow \psi_i ,\ 
	\tau \longrightarrow \xi, \ \varphi \longrightarrow N, 
	\lab{ghost1}
	\eeq
and 
	\beq
	\Lambda \longrightarrow  \phi 
	.\eeq
Ghosts in (\ref{ghost1}) are first generations, in particular, $c$ is 
Faddeev-Popov ghost, whereas $\phi$ is a second 
generation ghost. Their Grassmann parity and ghost number ($U$ number) 
are given by 
	\beq
	\begin{array}{ccccc}
	c &\ \psi_i &\ \xi &\ N &\ \phi \\
	1^-&\ 1^-&\ 1^-&\ 1^- &\ 2^+
	,\end{array}
	\lab{ghost2}
	\eeq
where the superscript of ghost number denotes the Grassmann parity. 
Note that the ghost number counts the degree of differential form on the 
moduli space ${\M}$ of the solution to the 3-d monopole equations. The 
minimal set $\Phi_{\mbox{\scriptsize min}}$ of fields consists of 
	\beq
	\begin{array}{ccccccccc}
	 A_i & \ b&\ M& \ G_i & \ \nu \\
	\ 0^+&\ 0^+&\ 0^+&\ 0^+&\ 0^+ ,
	\end{array}
	\eeq
and (\ref{ghost2}). 

On the other hand, the set of anti-fields $\Phi_{\mbox{\scriptsize min}}^*$ 
carrying opposite statistics to $\Phi_{\mbox{\scriptsize min}}$ is given by 
	\beq
	\begin{array}{ccccccccc}
	 A_{i}^*&\ b^*&\ M^*& \ G_i ^*& \ \nu^*& \ c^*&\ \psi_{i}^* 
	&\ N^* &\ \phi^* \\
	\ -1^-&\ -1^-&\ -1^-&\ -1^-&\ -1^-&\ -2^+&\ -2^+&\ -2^+&\ -3^-
	\end{array}
	.\eeq

Next step is to find a solution to the master equation 
with $\Phi_{\mbox{\scriptsize min}}$ and 
$\Phi_{\mbox{\scriptsize min}}^*$, given by  
	\beq
	\frac{\pa_r S}{\pa \Phi^A}\frac{\pa_l S}{\pa \Phi_{A}^*}-
	\frac{\pa_r S}{\pa \Phi_{A}^*}\frac{\pa_l S}{\pa \Phi^A}
	=0,\lab{redmaster}
	\eeq
where $r(l)$ denotes right (left) derivative.

The general solution for the first stage reducible 
theory at hand can be expressed by 
	\beqa
	S&=&S_c +\Phi_{i}^* R_{\alpha}^i C_{1}^{\alpha}+
	C_{1\alpha}^* (Z_{\beta}^{\alpha}C_{2}^{\beta}+
	T_{\beta\gamma}^{\alpha}C_{1}^{\gamma}C_{1}^{\beta})
	\nm \\
	& &+C_{2\gamma}^* A_{\beta\alpha}^{\gamma} 
	C_{1}^{\alpha}C_{2}^{\beta}
	+\Phi_{i}^* \Phi_{j}^* B_{\alpha}^{ji}C_{2}^{\alpha}+
	\cdots,
	\lab{solmas}
	\eeqa
where $C_{1}^{\alpha}(C_{2}^{\alpha})$ denotes generally the 
first (second) generation ghost and only relevant terms in our case 
are shown. We often use $\Phi_{\mbox{\scriptsize min}}^A =(\phi^i ,
C_{1}^{\alpha},C_{2}^{\beta})$, where $\phi^i$ denote generally 
the fields. In this expression, the indices should be interpreted as the 
label of fields. Do not confuse with space-time indices. 
The coefficients $Z_{\beta}^{\alpha}, T_{\beta\gamma}^{\alpha}$, 
etc can be directly determined from the master equation. In fact, 
it is known that these coefficients satisfy the following relations 
	\beqa
	R_{\alpha}^i Z_{\beta}^{\alpha}C_{2}^{\beta}-2\frac{\pa_r S_c}
	{\pa \phi^j}B_{\alpha}^{ji}C_{2}^{\alpha}(-1)^{|i|}&=&0,\nm \\
	\frac{\pa_r R_{\alpha}^i C_{1}^{\alpha}}{\pa \phi^j}
	R_{\beta}^j C_{1}^{\beta}+R_{\alpha}^i T_{\beta\gamma}^{\alpha}
	C_{1}^{\gamma}C_{1}^{\beta} &=&0,\nm \\
	\frac{\pa_r Z_{\beta}^{\alpha}C_{2}^{\beta}}{\pa \phi^j}
	R_{\gamma}^j C_{1}^{\gamma}+2T_{\beta\gamma}^{\alpha}
	C_{1}^{\gamma}Z_{\delta}^{\beta}C_{2}^{\delta}+
	Z_{\beta}^{\alpha}A_{\delta\gamma}^{\beta}
	C_{1}^{\gamma}C_{2}^{\delta}&=&0,
	\lab{coefficient}
	\eeqa
where $|i|$ means the Grassmann parity of the $i$-th field. 

In these expansion 
coefficients, $R_{\alpha}^i$ and $Z_{\beta}^{\alpha}$ are related to the local 
symmetry (\ref{alg}). On the other hand, as $T_{\beta\gamma}^{\alpha}$ is 
related to the structure constant of a given Lie algebra for a gauge 
theory, it is generally called as structure 
function. Of course if the theory is Abelian, such structure function 
does not appear. However, for a theory coupled with matters, 
all of the structure functions 
do not always vanish, even if the gauge group is Abelian. At first sight, this 
seems to be strange, but the expansion (\ref{solmas}) obviously 
detects the coupling of matter fields and ghosts. In fact, the appearance of 
this type of structure function is required in order to make the 
action to be constructed being full BRST invariant. 

After some algebraic works, we will find the solution to be 
	\beq
	S(\Phi_{\mbox{\scriptsize min}},\Phi_{\mbox{\scriptsize min}}^* )=
	S_c +\int_Y \Delta S d^3 x ,
	\eeq
where 
	\beqa
	\Delta S&=& A_{i}^* (\pa^i c +\psi^i ) +b^* \xi +M^* (icM+N) + 
	\ovM^* (-ic\ovM +\overline{N}) \nm \\
	& &+G_i ^* \left[\pa^{i}\xi -\epsilon^{ijk}\pa_j \psi_k +
	i(\overline{N}\sigma^{i0}M +\ovM \sigma^{i0}N)\right] \nm \\
	& &+\nu^* (ic\nu +i\gamma^{i}D_{i}N +\gamma^{i}\psi_{i}M+\gamma^0 bN+
	\gamma^0 \xi M)\nm\\
	& &+\overline{\nu}^* \overline{(ic\nu +i\gamma^{i}D_{i}N 
	+\gamma^{i}\psi_{i}M+\gamma^0 bN+\gamma^0 \xi M)}\nm \\
	& &+c^* \phi -\psi_{i}^* \pa^{i} \phi 
	-iN^* \left(\phi M +cN\right)+i\overline{N}^*
	 \left(\phi \ovM +c\ovN \right) \nm \\
	& &+2i\nu^* \overline{\nu}^* \phi 
	.\eeqa
	
We augment $\Phi_{\mbox{\scriptsize min}}$ by new fields $\chi_i ,
d_i ,\mu (\ovm),\zeta(\overline{\zeta}),\lambda,\rho,\eta,e$ 
and the corresponding anti-fields. 
Their ghost number and Grassmann patity are given by
	\beq
	\begin{array}{ccccccccc}
	\chi_i & \ d_i &\ \mu&\ \zeta&\ \lambda&\ \rho&\ \eta&\ e \\
	-1^-&\ 0^+&\ -1^-&\ 0^+&\ -2^+&\ -1^-&\ -1^-&\ 0^+
	\end{array} 
	\eeq
and 
	\beq
	\begin{array}{cccc}
	\chi_{i}^*&\ \mu^*&\ \lambda^*&\ \rho^*\\
	0^+&\ 0^+&\ 1^-&\ 0^+
	\end{array} 
	.\eeq
Then we look for the solution 
	\beq
	S'=S(\Phi_{\mbox{\scriptsize min}},\Phi_{\mbox{\scriptsize min}}^* )
	+\int_Y (\chi^{*i}d_{i}
	+\ovm^* \overline{\zeta}+\mu^* \zeta +\rho^* e +\lambda^* \eta)
	d^3 x  ,
	\lab{look}
	\eeq
where $d_{i},\zeta,e,\eta,$ are Lagrange 
multiplier fields. 

In order to obtain the quantum action we must fix the gauge. 
After a little thought, the best choice for the gauge fixing 
condition which can reproduce the action obtained from the dimensional 
reduction of the four dimensional one is found to be 
	\beqa
	G_i&=&0,\nm\\
	\nu &=&0,\nm\\
	\pa^i A_i  &=&0,\nm\\
	-\pa^{i}\psi_{i}+\frac{i}{2}(\ovN M-\ovM N)&=&0
	.\eeqa

Thus we can obtain the gauge fermion carrying the ghost number $-1$ and odd 
Grassmann parity, 
  	\beq
	\Psi =-\chi^{i}G_i -\ovm \nu -\mu \overline{\nu} 
	+\rho \pa^{i}A_{i} -
	\lambda\left[ -\pa^{i}\psi_{i}+\frac{i}{2}(\ovN M-\ovM N)\right]
	.\eeq
The quantum action $S_q$ can be obtained by eliminating anti-fields 
and are restricted to lie on the gauge surface  
	\beq
	\Phi^* =\frac{\pa_r \Psi}{\pa \Phi}.
	\eeq
Therefore the anti-fields will be 
	\beqa
	& &G_i ^* =-\chi_{i},\ \chi_{i}^* =-G_i,\ 
	\nu^* =-\ovm,\ \overline{\nu}^* =-\mu,\ \ovm^* =-\nu ,\ \mu^* 
	=-\ovn,\nm \\
	& &M^* =-\frac{i}{2}\lambda \ovN,\ \ovM^* =\frac{i}{2}\lambda  N,\ 
	N^* =\frac{i}{2} \lambda \ovM ,\ \ovN ^* =-\frac{i}{2} 
	\lambda M ,\nm \\
	& &\rho^* =\pa^{i}A_{i},\ A_{i}^* =-\pa_{i}\rho ,\ 
	\psi_{i}^* =-\pa_{i} \lambda ,\nm\\
	& &\lambda^* =-\left[-\pa^{i}\psi_{i} +
	\frac{i}{2}(\ovN M-\ovM N)\right],\ c^* =\phi^* =b^* =\zeta^* 
	(\overline{\zeta}^* )=0.
	\lab{surface1}
	\eeqa
Then the quantum action $S_q$ is given by  
	\beqa
	S_q &=&S'\left(\Phi,\Phi^* =\pa_r \Psi/\pa \Phi \right) ,
	\lab{quantum1}
	\eeqa
Substituting (\ref{surface1}) into (\ref{quantum1}), we find that 
	\beq
	S_q =S_c +\int_Y \widetilde{\Delta}Sd^3 x ,
	\eeq
where 
	\beqa
	\widetilde{\Delta}S&=&(-\triangle \phi +\phi \ovM M -i \ovN N)\lambda 
	-\left[-\pa^{i}\psi_{i}+\frac{i}{2}(\ovN M -\ovM N)
	\right] \eta \nm \\
	& &-\overline{\mu}(ic\nu +i\gamma^{i}D_{i}N +\gamma^{i}\psi_{i}M+
	\gamma^0 bN+\gamma^0 \xi M)\nm\\
	& &+\overline{(ic\nu +i\gamma^{i}D_{i}N +\gamma^{i}\psi_{i}M+
	\gamma^0 bN+\gamma^0 \xi M)}\mu+ 2i\phi \overline{\mu}\mu \nm \\
	& &-\chi^{i} 
	\left[\pa_{i}\xi -\epsilon_{ijk}\pa^j \psi^{k} +i(\overline{N} 
	\sigma_{i0}M +\ovM \sigma_{i0}N)\right] \nm\\
	& &+\rho (\triangle c +\pa^{i}\psi_{i})-d^{i}
	G_i -\overline{\zeta}\nu -\ovn \zeta +e \pa^{i}A_i  
	.\lab{act1}
	\eeqa
Using the condition (\ref{con}) with $c=0$, we can arrive at 
	\beq
	S_{q}'=S_c |_{G_i =\nu (\ovn )=0}+\int_Y \widetilde{\Delta}S|_{c=0} d^3 x
	,\lab{act111}
	\eeq
where
	\beqa
	\widetilde{\Delta}S|_{c=0}&=&(-\triangle \phi +\phi \ovM M -i \ovN N)\lambda 
	-\left[-\pa^{i}\psi_{i}+\frac{i}{2}(\ovN M -\ovM N)
	\right] \eta \nm \\
	& &-\overline{\mu}(i\gamma^{i}D_{i}N +\gamma^{i}\psi_{i}M
	+\gamma^0 bN+\gamma^0 \xi M)\nm\\
	& &+\overline{(i\gamma^{i}D_{i}N +\gamma^{i}\psi_{i}M+\gamma^0 bN+
	\gamma^0 \xi M)}\mu+ 2i\phi \overline{\mu}\mu \nm \\
	& &-\chi^{i} 
	\left[\pa_{i}\xi-\epsilon_{ijk}\pa^j \psi^{k} +i(\overline{N} 
	\sigma_{i0}M +\ovM \sigma_{i0}N)\right] \nm\\
	& &+\rho \pa^{i}\psi_{i} +e \pa^{i}A_i
	.\eeqa
It is easy to find that (\ref{act111}) is consistent with the action 
found by the dimensional reduction of the four dimensional 
topological action \cite{ZWCM}. 

\begin{center}
\subsubsection{BRST transformation}
\end{center}

The Batalin-Vilkovisky algorithm also facilitates to construct BRST 
transformation rule. The BRST transformation rule for a field $\Phi$ is 
defined by 
	\beq
	\delta_B \Phi =\epsilon \left.\frac{\pa_r S'}{\pa \Phi^*}
	\right|_{\Phi^* =\frac{\pa_r \Psi}{\pa \Phi}},
	\lab{br}
	\eeq
where $\epsilon$ is a constant Grassmann odd parameter. With this 
definition for (\ref{act1}), we obtain 
	\beqa
	\delta_B A_{i} &=&-\epsilon(\pa_{i}c +\psi_{i}) , \nm \\
	\delta_B b&=&-\epsilon \xi,\nm\\
	\delta_B M&=&-\epsilon(icM +N), \nm \\
	\delta_B G_i &=&-\epsilon \left[\pa_{i}\xi -\epsilon_{ijk}\pa^j 
	\psi^{k}+i(\ovN\sigma_{i0}M+\ovM \sigma_{i0}N)
	\right], \nm\\
	\delta_B \nu&=&-\epsilon(ic\nu +i\gamma^{i}D_{i}N +\gamma^{i}\psi_{i}M
	+\gamma^0 bN+\gamma^0 \xi M-i\mu\phi),\nm \\
	\delta_B c&=&\epsilon \phi ,\nm \\
	\delta_B \psi_i &=&-\epsilon \pa_i \phi,\nm \\
	\delta_B \rho &=& \epsilon e,\nm \\
	\delta_B \lambda &=& -\epsilon \eta,\nm\\
	\delta_B \mu &=& \epsilon \zeta,\nm\\
	\delta_B N &=& -i\epsilon (\phi M+cN),\nm\\
	\delta_B \chi_{i}&=&\epsilon d_{i},\nm\\
	\delta_B \phi &=&\delta_B \xi =\delta_B d_{i}=
	\delta_B e =\delta_B \zeta =\delta_B \eta =0 .
	\lab{BRST1}
	\eeqa
It is clear at this stage that (\ref{BRST1}) has on-shell nilpotency, i.e., 
the quantum equation of motion for $\nu$ must be used in order to 
have $\delta_{B}^2 =0$. This is due to the fact that the gauge algebra has 
on-shell reducibility. Accordingly, the Batalin-Vilkovisky 
algorithm gives a BRST invariant action and on-shell nilpotent BRST 
transformation. Note that the equations 
	\beqa
	\pa_{i}\xi-\epsilon_{ijk}\pa^j \psi^{k}+
	i(\ovN\sigma_{i0}M+\ovM \sigma_{i0}N)&=&0,\nm\\
	i\gamma^{i}D_{i}N +\gamma^{i}\psi_{i}M
	+\gamma^0 bN+\gamma^0 \xi M&=&0
	\eeqa
can be recognized as linearizations of the 3-d monopole equations and the 
number of linearly independent solutions gives the dimension of ${\M}$. 
 
It is now easy to show that the global supersymmetry can be recovered 
from (\ref{BRST1}). In Witten type theory, $Q_B$ can be interpreted as a 
supersymmetric BRST charge. 
We define the supersymmetry transformation as 
	\beq
	\delta_S \Phi :=\delta_B \Phi |_{c=0}.
	\eeq
We can easily find that the result is consistent with the supersymmetry 
algebra of Ref. 28.
	
\begin{center}
\subsubsection{Off-shell action}
\end{center}
 
As was mentioned before, the quantum action of Witten type 
topological field theory can be represented by BRST commutator 
with nilpotent BRST charge $Q_B$. However, since our 
BRST transformation rule is on-shell nilpotent, we should 
integrate out $\nu$ and $G_i$ in order to obtain off-shell BRST 
transformation and off-shell quantum action. 

For this purpose, let us consider the following terms in (\ref{act1}),
	\beq
	\frac{1}{2}(G_i -X_i )^2+\frac{1}{2}|\nu -A|^2 -i\ovm c\nu +\overline{
	ic\nu}\mu -\overline{\zeta}\nu -\ovn \zeta-d^i G_i,
	\lab{arra1}
	\eeq
where 
	\beq
	X_i =\pa_i b -\frac{1}{2}\epsilon_{ijk}F^{jk}+i\ovM 
	\sigma_{i0}M,\ A=i\gamma^i D_i M+\gamma^0 bM.
	\eeq
Here, let us define  
	\beq
	\nu'=\nu -A,\ B=-ic\mu-\zeta
	.\eeq
$\nu'(\ovn')$ and $G_i$ can be integrated out and then (\ref{arra1}) will be  
	\beq
	-\frac{1}{2}d_i d^i -d_i X^i -2|B|^2 +\overline{B}A +B\overline{A}
	.\eeq 
Consequently, we obtain the off-shell quantum action 
	\beq
	S_q =\left\{ Q,\widetilde{\Psi}\right\}
	,\lab{result1}
	\eeq
where 
	\beqa
	\widetilde{\Psi}&=&
	-\chi^{i}\left(X_i +\frac{\alpha}{2}d_i \right) -\mu (\overline{i\gamma^i 
	D_i M +\gamma^0 b M -\beta B}) -\ovm (i\gamma^i 
	D_i M +\gamma^0 b M -\beta B)+\rho\pa^i A_i \nm\\
	& &-\lambda\left[-\pa^i \psi_i +
	\frac{i}{2}(\ovN M-\ovM N)\right].
	\eeqa
$\alpha$ and $\beta$ are arbitrary gauge fixing paramaters. Convenience choice for 
them is $\alpha=\beta=1$. The BRST transformation 
rule for $X_i$ and $B$ fields can be easily obtained, although we do not 
write down here.

\begin{center}
\subsubsection{Observables}
\end{center}

We can now discuss the observables. For this purpose, let us define \cite{BG}  
	\beqa
	{\A}&=&A+c,\nm\\
	{\F}&=&F+\psi-\phi,\nm\\
	{\K}&=&db+\xi ,
	\eeqa
where we have introduced differential form notations, but their meanings would 
be obvious. $A$ and $c$ are considered as a $(1,0)$ and $(0,1)$ part of 
1-form on $(Y,{\M})$. Similarly, $F,\psi$ and $\phi$ are $(2,0),(1,1)$ 
and $(0,2)$ part 
of the 2-form ${\F}$, and $db$ and $\xi$ are $(1,0)$ and $(0,1)$ part of the 
1-form ${\K}$. Thus ${\A}$ defines a connection 1-form on $(Y,{\M})$ 
and ${\F}$ is a curvature 2-form. Note that the exterior derivative $d$ 
maps any $(p_1 ,p_2)$-form 
$X_p$ of total degree $p=p_1 +p_2$ to $(p_1 +1,p_2 )$-form, 
but $\delta_B$ maps any $(p_1 ,p_2)$-form to $(p_1 ,p_2 +1)$-form. Also 
note that 
	\beq
	X_p X_q =(-1)^{pq} X_q X_p 
	.\eeq 
Then the action of $\delta_B$ is 
	\beqa
	(d+\delta_B ){\A}&=&{\F},\nm\\
	(d+\delta_B )b &=&{\K}
	.\lab{hosi}
	\eeqa	
${\F}$ and ${\K}$ also satisfy 
	\beqa
	(d+\delta_B ){\F} &=&0,\nm\\
	(d+\delta_B ){\K}&=&0
	.\lab{Bianchi}
	\eeqa
Eq.(\ref{Bianchi}) can be interpreted as Bianchi identities in 
Abelian theory. Eqs.(\ref{hosi}) and 
(\ref{Bianchi}) mean anti-commuting property between the BRST 
variation $\delta_B$ and the exterior differential $d$, 
i.e., $\{\delta_B ,d\}=0$. 

The BRST transformation rule in geometric sector can be easily read 
from (\ref{BRST1}), i.e., $\delta_B A,\delta_B \psi,\delta_B c$ and 
$\delta_B \phi$. Eq. (\ref{Bianchi}) implies 
	\beq
	(d+\delta_B ) {\F}^n =0
	\lab{nth}
	\eeq
and expanding the above expression by ghost number and form degree, we 
obtain the following $(i,2n-i)$-form $W_{n,i}$,
	\beqa
	W_{n,0}&=&\frac{\phi^n}{n!} ,\nm\\
	W_{n,1}&=&\frac{\phi^{n-1}}{(n-1)!}\psi,\nm\\
	W_{n,2}&=&\frac{\phi^{n-2}}{
	2(n-2)!}\psi \wedge \psi -\frac{\phi^{n-1}}{(n-1)!}F,\nm\\
	W_{n,3}&=& \frac{\phi^{n-3}}{6(n-3)!}\psi \wedge \psi \wedge \psi  
	-\frac{\phi^{n-2}}{(n-2)!}F\wedge \psi,\lab{obco}
	\eeqa
where 
	\beqa
	0&=& \delta_B W_{n,0},\nm\\
	d W_{n,0} &=& \delta_B W_{n,1},\nm\\
	d W_{n,1} &=&\delta_B W_{n,2},\nm\\
	d W_{n,2} &=&\delta_B W_{n,3},\nm\\
	d W_{n,3} &=& 0.
	\lab{ob3}
	\eeqa
 
Picking a certain $k$-cycle $\gamma$ as a representative and defining the 
integral 
	\beq
	W_{n,k}(\gamma)=\int_{\gamma} W_{n,k},
	\eeq
we can easily prove 
	\beqa
	\delta_B W_{n,k}(\gamma)&=&-\int_{\gamma} dW_{n,k-1}\nm\\
	&=&-\int_{\pa \gamma} W_{n,k-1}\nm\\
	&=&0,
	\eeqa
as a consequence of (\ref{ob3}). Note that the last equality follows 
from the fact that the cycle $\gamma$ is a simplex without boundary, i.e., 
$\pa \gamma =0$. Therefore, $W_{n,k}(\gamma)$ indeed gives a topological 
invariant associated with $n$-th Chern class on $Y\times {\M}$. 

On the other hand, since we have a scalar field $b$ and its ghosts, 
we may construct topological observables 
associated with them. Therefore, 
the observables can be obtained from the ghost expansion of 
	\beq
	(d+\delta_B ){\F}^n \wedge {\K}^m =0
	.\eeq
Explicitly, for $m=1$, for example, we obtain  	
	\beqa
	0&=& \delta_B W_{n,1,0},\nm\\
	d W_{n,1,0} &=&\delta_B W_{n,1,1} ,\nm\\
	d W_{n,1,1}&=&\delta_B W_{n,1,2},\nm\\
	d W_{n,1,2}&=&\delta_B W_{n,1,3},\nm\\
	d W_{n,1,3}&=&0,
	\eeqa
where
	\beqa
	 W_{n,1,0}&=&\frac{\phi^n}{n!}\xi,\nm\\
	 W_{n,1,1} &=&\frac{\phi^{n-1}}{(n-1)!}\psi\xi -
	\frac{\phi^n}{n!}db,\nm\\
	 W_{n,1,2} &=&\frac{\phi^{n-2}}{2(n-2)!}
	\psi \wedge \psi \xi -\frac{\phi^{n-1}}{(n-1)!}F\xi 
	-\frac{\phi^{n-1}}{(n-1)!} \psi \wedge db,\nm\\
	W_{n,1,3}&=&\frac{\phi^{n-3}}{6(n-3)!}\psi\wedge\psi\wedge\psi \xi +
	\frac{\phi^{n-1}}{(n-1)!}F\wedge db\nm\\
	& &+\frac{\phi^{n-2}}{2(n-2)!}(2\psi\wedge F\xi +\psi\wedge\psi
	\wedge db) 
	.\lab{57}
	\eeqa
These corresponds to the cocycles \cite{BG} in $U(1)$ case.

Next, let us look for the observables for matter sector. The BRST 
transformation rules in this sector is given by 
$\delta_B ,\delta_B N,\delta_B c$ and $\delta_B \phi$. 
At first sight, the matter sector does not have any observable, but we can 
find the combined form 
	\beq
	\widetilde{W}=i\phi \ovM M+\ovN N
	\lab{comb}
	\eeq
is an observable. However, unfortunately, as $\widetilde{W}$ is 
cohomologically trivial because $\delta_B \widetilde{W}=0$ but 
$d\widetilde{W} \neq \delta_B \widetilde{W}'$ for some $\widetilde{W}'$. 
Accordingly, $\widetilde{W}$ does not give any new topological invariant. 
Hyun {\em et al.} \cite{HPP1,HPP2} identified $\widetilde{W}$ as a part of 
the bare mass term of the hypermultiplet in their twisting construction of 
topological QCD in four dimensions. 

In topological Bogomol'nyi theory, there is a sequence of observables 
associated with a magnetic charge. For the Abelian case, it is given by 
	\beq
	W=\int_Y F \wedge db 
	.\lab{charge}
	\eeq
As is pointed out for the case of Bogomol'nyi monopoles \cite{BRT}, 
we can not obtain the observables related with 
this magnetic charge by the action of $\delta_B$ as well, but we can 
construct those observables by anti-BRST variation $\overline{\delta}_B$ 
which maps $(m,n)$-form to $(m,n-1)$-form. 
$\overline{\delta}_B$ can be obtained by a discrete symmetry which is realized 
as ``time reversal symmetry'' in four dimensions. In our three dimensional 
theory, the discrete symmetry is given by 
	\beqa
	& & \phi \longrightarrow -\lambda,\ \lambda \longrightarrow -\phi,\ 
	N \longrightarrow i\sqrt{2}\mu,\ \mu 
	\longrightarrow\frac{i}{\sqrt{2}}N ,\nm\\
	& &\psi_i \longrightarrow \frac{\chi_i}{\sqrt{2}},\ \chi_i 
	\longrightarrow 
	\sqrt{2}\psi_i ,\ \eta \longrightarrow \sqrt{2}\xi,\ \xi 
	\longrightarrow -
	\frac{\eta}{\sqrt{2}} 
	\lab{inv1}
	\eeqa
with 
	\beq
	b \longrightarrow -b
	.\lab{inv2}
	\eeq
Eq.(\ref{inv2}) is an additional symmetry \cite{BRT}. Note 
that we must also change $N$ and $\mu$ (and their conjugates). The positive 
chirality condition for $M$ should be used in order to check the invariance 
of the action. In this way, we can obtain 
anti-BRST transformation rule by substituting (\ref{inv1}) and (\ref{inv2}) 
into (\ref{BRST1}) and then we can obtain the observables associated with 
the magnetic charge by using the action of this anti-BRST 
variation \cite{BRT}. 

The topological observables available in this theory are the same with 
those of topological Bogomol'nyi monopoles.  
 
Finally, let us briefly comment on our three 
dimensional theory. First note that Lagrangian $L$ and Hamiltonian $H$ 
in dimensional reduction can be considered as equivalent. 
This is because the relation between them is defined by 
	\beq
	H=p \dot{q} -L ,
	\eeq
where $q$ is any field, the overdot means time derivative 
and $p$ is a canonical conjugate momentum of $q$, and the dimensional 
reduction requires the time independence of all fields, thus $H =-L$ in 
this sense. Though we have constructed the three dimensional 
action directly from the 3-d monopole equations, our action 
may be interpreted essentially as the Hamiltonian of the four 
dimensional Seiberg-Witten theory. In this sense \cite{W1}, 
the ground states may correspond to the ``Floer groups ?'' of $Y$, 
but we do not know the precise correspondence. 

\begin{center}
\subsection{Non-Abelian case}
\end{center}

It is easy to extend the results obtained in the previous 
subsection to non-Abelian case. In this subsection, we summarize the results 
for the non-Abelian 3-d monopoles. 

\begin{center}
\subsubsection{Non-Abelian topological action}
\end{center}

With the auxiliary fields $G_{\mu\nu}^a$ and $\nu$, we consider  
	\beq
	S_c =\frac{1}{2}\int_Y d^3 x \left[ 
	(G_i ^{a} -K_{i}^a )^2 +
	|\nu -i\gamma^{i}D_i M -\gamma^0 b M|^2 \right] , 
	\lab{startt}
	\eeq
where 
	\beq
	K_{i}^a =\pa_i b^a +f_{abc}A_{i}^b b^c -\frac{1}{2}\epsilon_{ijk}
	F_{jk}^a +i\ovM \sigma_{i0}T^a M
	.\eeq
Note that the minimum of (\ref{startt}) with the gauge 
	\beq
	G_i ^a =\nu =0
	\eeq
are given by the non-Abelian 3-d monopoles. We take 
the generator of Lie algebra in the fundamental representation, 
e.g., for $SU(n)$,  
	\beq
	(T_a )_{ij}(T^a )_{kl}=\delta_{il}\delta_{jk}-\frac{1}{n}
	\delta_{ij}\delta_{kl}.
	\eeq
Extension to other Lie algebra and representation is straightforward. 

The gauge transformation rule for (\ref{startt}) is given by  
	\beqa
	\delta A_{i}^a &=&\pa_{i} \theta^a +f_{abc}A_{i}^b \theta^c 
	+\epsilon_{i}^a , \nm \\
	\delta b^a &=&f_{abc}b^b \theta^c +\tau^a ,\nm\\
	\delta M&=&i\theta M +\varphi, \nm \\
	\delta G_i ^a &=&f_{abc}G_i ^b \theta^c +\left[-\epsilon_{ijk}(
	\pa^{j}\epsilon^{ak} +f_{abc}\epsilon^{jb}A^{ck}) \right.\nm\\
	& &+\left. \pa_i \tau^a +f_{abc}(\epsilon_{i}^b b^c 
	-\tau^b A_{i}^c )+i(\overline{\varphi}\sigma_{i0}T^a M+\ovM \sigma_{i0}
	T^a \varphi )\right], \nm\\
	\delta \nu&=& i\gamma^{i}D_{i}\varphi +\gamma^{i}\epsilon_{i} M 
	+\gamma^0 b \varphi +\gamma^0 \tau M +i\theta \nu .
	\lab{gaugealgg}
	\eeqa
Note that we have a $G_i ^a$ term 
in the transformation of $G_i ^a$, while it did not appear in 
Abelian theory. 

The gauge algebra (\ref{gaugealgg}) 
possesses on-shell zero modes as in the Abelian case. Setting 
	\beq
	\theta^a =\Lambda^a,\ \epsilon_{i}^a =-\pa_{i}\Lambda^a -
	f_{abc}A_{i}^b \Lambda^c,\ 
	\tau^a =-f_{abc}b^b \Lambda^c ,\ \varphi =-i\Lambda M,
	\eeq	
we can easily find that (\ref{gaugealgg}) closes
	\beqa
	\delta A_{i}^a &=&0, \nm \\
	\delta b^a &=&0,\nm\\
	\delta M& =&0, \nm \\
	\delta G_i^a &=& f_{abc}\Lambda^c [G_i^b -K_{i}^b ]
	|_{\mbox{\scriptsize on\mbox{-}shell}}=0,\nm\\
	\delta\nu &=& i\Lambda[\nu-i(\gamma^i D_i -i\gamma^0 b)M]|_{
	\mbox{\scriptsize on\mbox{-}shell}} =0 , 
	\lab{gaugealgg2}
	\eeqa
when the equations of motion of $G_i ^a$ and $\nu$ are used.
Note that we must use both equations of motion of $G_i ^a$ and 
$\nu$ in the non-Abelian case, while only ``$\nu$'' was 
needed for the Abelian theory. Furthermore, as $\varphi$ is a 
parameter in the spinor space, $\varphi$ is not {\bf g}-valued, 
in other words, $\varphi \neq \varphi^a T^a$. 
Eq.(\ref{gaugealgg}) is first stage reducible.  

The assortment of ghost fields, the minimal 
set $\Phi_{\mbox{\scriptsize min}}$ of the fields and the ghost 
number and the Grassmann parity, furthermore those for 
$\Phi_{\mbox{\scriptsize min}}^*$ would be obvious. 

Then the solution to the master equation will be 
	\beq
	S(\Phi_{\mbox{\scriptsize min}},\Phi_{\mbox{\scriptsize min}}^* )=
	 S_c +\int_Y \mbox{tr}\ \Delta S_n d^3 x 
	,\eeq
where 
	\beqa
	\Delta S_n &=&
	A_{i}^{*}(D^{i}c+\psi^{i})+b^* (i[b,c]+\xi )\nm\\
	& &+M^* (icM+N)+\ovM^* (-ic\ovM +\ovN ) \nm\\
	& &+G_i ^{*}\widetilde{G}^{i}
	-iN^{*} (\phi M+cN)+i\ovN^{*}(\phi\ovM+c\ovN)\nm \\
	& &+\nu^* (ic\nu +i\gamma^{i}D_{i}N 
	+\gamma^{i}\psi_{i}M+\gamma^0 bN+\gamma^0 \xi M)\nm\\
	& &+\overline{\nu}^* 
	\overline{(ic\nu +i\gamma^{i}D_{i}N 
	+\gamma^{i}\psi_{i}M+\gamma^0 bN+\gamma^0 \xi M)}\nm\\
	& &+2i\nu^* \ovn^* \phi
	+\psi_{i}^{*}(-D^{i}\phi-i\{\psi^i ,c\}) 
	+c^{*}\left(\phi-\frac{i}{2}\{c,c\}\right)-i\phi^{*}[\phi,c] 
	\nm\\
	 & &-\frac{i}{2}\{G_i ^{*},G ^{*i}\}\phi+i\xi^* ([b,\phi]-\{\xi,c\}) 
	.\eeqa
Here
	\beq
	\widetilde{G}_{i}=i[c,G_i ]- 
	\epsilon_{ijk}D^j \psi^k +D_i \xi +[\psi_i ,\xi]+i(\overline{N}
	\sigma_{i0}T_a T^a M+\ovM \sigma_{i0}T_a T^a N)
	.\eeq
The equations 
	\beqa
	-\epsilon_{ijk}D^j \psi^k +D_i \xi +[\psi_i ,\xi]
	+i(\overline{N}\sigma_{i0}T_a T^a M+\ovM \sigma_{i0}
	T_a T^a N) 
	&=&0,\nm\\
	i\gamma^{i}D_{i}N+\gamma^{i}\psi_{i}M +\gamma^0 bN+\gamma^0 \xi M&=&0,
	\eeqa
can be seen as linearizations of non-Abelian 3-d monopoles. 

We augment $\Phi_{\mbox{\scriptsize min}}$ by new fields $\chi_{i}^a ,
d_{i}^a ,\mu (\ovm),\zeta(\overline{\zeta}),\lambda,\rho,\eta,e$ 
and the corresponding anti-fields, but Lagrange multipliers fields 
$d_{i}^a ,\zeta(\overline{\zeta}),e,\eta,$ are 
assumed not to have anti-fields for simplicity and therefore their 
BRST transformation rules are set to zero. This simplification means that we 
do not take into account of BRST exact terms. In this sense, the result to be 
obtained will correspond to those of the dimensionally reduced version 
of the four dimensional theory \cite{LM2,LM3,HPP1,HPP2} up to 
these terms, i.e., topological numbers. 

From the gauge fixing condition 
	\beqa
	G_i ^a &=&0,\nm\\
	\nu &=&0,\nm\\
	\pa^{i}A_{i} &=&0,\nm\\
	-D^{i}\psi_{i}+\frac{i}{2}(\ovN M-\ovM N)&=&0
	,\eeqa
the gauge fermion will be   
	\beq
	\Psi =-\chi^{i}G_i -\ovm \nu -\mu \overline{\nu} 
	+\rho \pa^{i}A_{i} -\lambda\left[ -D^{i}\psi_{i}
	+\frac{i}{2}(\ovN M-\ovM N)\right].
	\eeq

The anti-fields are then given by  
	\beqa
	& &G_i ^* =-\chi_{i},\ \chi_{i}^* =-G_i ,\ 
	\nu^* =-\ovm,\ \overline{\nu}^* =-\mu,\ \ovm^* =-\nu ,\ \mu=-\ovn,
	\nm \\
	& &M^* =-\frac{i}{2}\lambda \ovN,\ \ovM^* =\frac{i}{2}\lambda  N,\ 
	N^* =\frac{i}{2} \lambda \ovM ,\ \ovN ^* =-\frac{i}{2} \lambda M ,
	\nm \\
	& &\rho^* =\pa^{i}A_{i},\ A_{i}^* =-\pa_{i}\rho
	+i[\lambda,\psi_{i}],\ \psi_{i}^* =-D_{i} \lambda ,\nm\\
	& &\lambda^* =-\left[-D_{i}\psi^{i}+[b,\xi]
	+\frac{i}{2}(\ovN M-\ovM N)\right],\ b^* =c^*=\xi^* =\phi^* =\zeta^*  
	(\overline{\zeta}^* )=0
	.\eeqa
Therefore we find the quantum action
	\beq
	S_q =S_c +\int_Y \mbox{tr}\ \widetilde{\Delta}S_n d^3 x
	,\eeq
where 
	\beqa
	\widetilde{\Delta}S_n &=&-\left[-D_i \psi^i +[b,\xi] +\frac{i}{2}(
	\ovN M -\ovM N)\right] \eta
	-\lambda(D_i D^i \phi +iD_i \{ \psi^i ,c\})\nm\\
	& &+i\lambda\{\psi_i ,D^i c+\psi^i \}+
	(\phi\ovM M -i\ovN N)\lambda \nm\\
	& &-\chi^{i}\left[i[c,G_i ]+
	\epsilon_{ijk}D^j \psi^k +D_k \xi +[\psi_k ,\xi]
	+\frac{i}{2}(\ovN\sigma^{ij}T_a T^a M +
	\ovM \sigma^{ij}T_a T^a N)\right] \nm \\
	& &-\ovm (i\gamma^{\mu}D_{\mu}N+\gamma^{\mu}\psi_{\mu}M+ic\nu)
	+\overline{(i\gamma^{i}D_{i}N+\gamma^{\mu}\psi_{\mu}M+ic\nu)}\mu \nm\\
	& &+2i\phi \ovm\mu -\frac{i}{2}\{\chi_{i},\chi^{i}\}\phi +
	\rho(\pa_i D^i c+\pa_{i}\psi^{i})\nm \\
	& &-d^{i}G_i -\overline{\zeta}\nu -\ovn \zeta +e\pa^{i}A_{i} 
	.\lab{act}
	\eeqa
In this quantum action, setting 
	\beq
	M(\ovM)=N(\ovN)=\mu (\ovm)=\nu(\overline{\nu})=0,
	\eeq	
we can find that the resulting action coincides with 
that of Bogomol'nyi monopoles \cite{BRT}. 

Finally, in order to obtain the off-shell quantum action, both the 
auxiliary fields should be integrated out by the similar 
technique presented in Abelian case, but we remain it the reader's 
exercise. 

\begin{center}
\subsubsection{BRST transformation}
\end{center}

The BRST transformation rule is given by 
	\beqa
	\delta_B A_i&=&-\epsilon(D_{i}c+\psi_{i}),\nm\\
	\delta_B b&=& -\epsilon (i[c,b]+\xi),\nm\\
	\delta_B \xi &=&i\epsilon ([b,\phi ]-\{ \xi,c \} ),\nm\\
	\delta_B M&=&-\epsilon(icM +N), \nm \\
	\delta_B G_i&=&-\epsilon(\widetilde{G}_i -i[\chi_i ,\phi]), \nm\\
	\delta_B \nu&=&-\epsilon(ic\nu +\gamma^{\mu}D_{\mu}N+\gamma^{\mu}
	\psi_{\mu} M -i\mu\phi ) ,\nm\\
	\delta_B c&=&\epsilon \left(\phi-\frac{i}{2}\{c,c\}\right),\nm \\
	\delta_B \psi_{i}&=&-\epsilon \left(D_{i}\phi+i\{\psi_i ,c\}\right),
	\nm \\
	\delta_B \rho &=& \epsilon e,\nm \\
	\delta_B \lambda &=&-\epsilon \eta,\nm\\
	\delta_B \mu &=& \epsilon \zeta,\nm\\
	\delta_B N &=&-i\epsilon (\phi M+cN),\nm\\
	\delta_B \chi_{i}&=&\epsilon d_{i},\nm\\
	\delta_B \phi &=&i\epsilon [\phi,c],\nm\\
	\delta_B d_{i}&=&\delta_B e =\delta_B \zeta=\delta_B \eta =0 .
	\lab{BRST}
	\eeqa

It is easy to obtain supersymmetry also in this case. However, as we 
have omitted the BRST exact terms, the supersymmetry in our construction 
does not detect them.

\begin{center}
\subsubsection{Observables}
\end{center}

We have already constructed the topological observables for Abelian case. 
Also in non-Abelian case, the construction of observables is basically 
the same. But the relation (\ref{hosi}) and (\ref{Bianchi}) are 
required to modify 
	\beqa
	(d+\delta_B ){\A}-\frac{i}{2}[{\A},{\A}]&=&{\F},\nm\\
	(d+\delta_B )b -i[{\A},b] &=&{\K}
	\lab{78}
	\eeqa
and 
	\beqa
	(d+\delta_B ){\F}-i[{\A},{\F}]&=&0,\nm\\
	(d+\delta_B ){\K}-i[{\A},{\K}]&=&i [{\F},b]
	,\lab{79}
	\eeqa
respectively, where $[*,*]$ is a graded commutator. 
The observables in geometric and matter sector are the same as 
before, but we should replace $db$ by $d_A b$ in (\ref{57}) as well as 
(\ref{78}) and (\ref{79}), 
where $d_A$ is a exterior covariant derivative and trace is required. 
In addition, 
the magnetic charge observables are again obtained by 
anti-BRST variation as outlined berfore. 

The observables in geometric sector are those in (\ref{obco}) and follow 
the cohomological relation (\ref{ob3}). In this way, the topological 
observables available in this three dimensional theory are precisely 
the Bogomol'nyi monopole cocycles \cite{BG}.

\begin{center}
\section{Summary}
\end{center}

\renewcommand{\theequation}{5.\arabic{equation}}\setcounter{equation}{0}

We have presented the existence of the topological field theories 
which describe the moduli space of Abelian and non-Abelian three dimensional 
Seiberg-Witten monopole equations by using the Batalin-Vilkovisky 
quantization procedure. In the Abelian case, our topological action with a 
certain gauge condition is found to be consistent with that of the 
dimensionally reduced version of the four dimensional one. 
We have also established the three dimensional 
non-Abelian action. The interesting point is that this non-Abelian action 
can be viewed as the Bogomol'nyi monopole topological action including 
matter and its associated ghost. We have easily obtained the BRST and 
anti-BRST transformation rules. The topological observables related to the 
Chern 
classes can be found by the standard fashion. We have found that they are 
precisely the cocycles of Bogomol'nyi monopole topological field theory. 

In this paper, we have not include the mass term for the Weyl spinor, but the 
introduction of the mass term may connect the Bogomol'nyi and the 3-d 
Seiberg-Witten monopole topological field theory, as was shown 
that the mass term interporates Donaldson theory and Seiberg-Witten theory 
in four dimensions \cite{HPP2}. This point of view should be further studied.

We have already known various progress on the self-dual Yang-Mills equation, 
but there remain several tasks for Seiberg-Witten equations, so let us briefly 
comment on them as open problems. 

\begin{enumerate}
\item
Integrability of Seiberg-Witten equations. 

As is well-known, the self-dual Yang-Mills equation can be reduced to some 
solitonic equations such as non-linear 
Schr\"{o}dinger equation or KdV equation after suitable choice for the gauge 
fields \cite{MS,Str}, although there is no proof that the self-dual 
Yang-Mills equation is indeed integrable. 
On the other hand, as for the Seiberg-Witten equations, they can not be 
viewed as integrable equations at first sight, but it was found that the 
Seiberg-Witten equations on $\RR$ could be realized as Liouville vortex 
equations which are manifestly integrable \cite{NS} (as for a solution on 
$\RRR$, there is a Freund's solution \cite{F}). This fact seems to 
connect integrable systems and Seiberg-Witten monopoles, but unfortunately we 
do not know whether there exist another examples of integrable systems 
related to the Seiberg-Witten monopoles. Furthermore, as explicit solutions 
to non-Abelian Seiberg-Witten equations have not ever been found, we can not 
pursue the integrability. For this 
direction, twistor program \cite{PM} may be available, as is often used for 
the self-dual Yang-Mills equation \cite{MS,Ward,CG}. 

\item
Reduction to two dimensional surfaces (Riemann surfaces $\Sigma$). 

We can dimensionally reduce the Seiberg-Witten equations onto 
two dimensional surfaces. 
As have been mentioned before, the operation of dimensional reduction 
connects the theories between four and three, and three and two dimensions, 
the two dimesional theory may be regarded as a dual $U(1)$ theory for the 
$SU(2)$ Hitchin equations \cite{Hit} (usually, the Lie group for 
Hitchin equations is taken $SO(3)$ rather than $SU(2)$), 
i.e., two dimensional Yang-Mills-Higgs equations. 
One approach to study this observation is to construct solutions to the 
Seiberg-Witten equations on Riemann surfaces and compare their properties 
with those of the Hitchin equations. Recently, the 
reduced Seiberg-Witten equations were studied and it was pointed out that 
the set of equations had an extremely similar structure to the Hitchin 
equations, except the distinction of Higgs field and 
Weyl spinor \cite{MR}. We would like to interprete the relationship between 
these two theories in the context of topological quantum field theory, but 
any progress such as the study of the topological field 
theory associated with the two dimensional Seiberg-Witten monopoles 
has not ever been made (a topological action is obtained by a 
dimensional reduction \cite{O}), although Yang-Mills theory on Riemann 
surfaces are well discussed (see e.g., Ref. 2 and references therein). 
We know that the Yang-Mills-Higgs theory in two dimensions is 
closely related to a conformal field theory \cite{CGG}, but is it true 
also in two dimensional Seiberg-Witten theory?
\end{enumerate}

There are other problems such as supersymmetric extension \cite{AG,AGW}, 
twistor description \cite{V,NSS}, 
and so on, but a lot of (topological) field theoretical 
techniques to study these problems have been developed. Nevertheless, 
the Seiberg-Witten theory does not seem to be fully discussed even in 
four dimensions as 
well as lower dimensions in contrast with the Donaldson theory in 
the context of topological quantum field theory. Filling the gap may be 
attract problem, but much efforts will be required. 

\begin{center}
\section*{Acknowledgement}
\end{center}

I am grateful to Dr. H. Kanno for helpful suggestions and comments on 
the manuscript.

\begin{center}
\section*{Appendix A. Non-Abelian Seiberg-Witten theory}
\end{center}

\renewcommand{\theequation}{A\arabic{equation}}\setcounter{equation}{0}

In this Appendix, we summarize on 
the construction of the topological action for non-Abelian Seiberg-Witten 
monopoles in four dimensions by Batalin-Vilkovisky algorithm. 

Let $P$ be a principal bundle with a compact simple Lie group $G$ and 
by $E$ we mean the associated bundle to $P$. Then the gauge field is the 
connection on $E$ and $M$ is a section of $S^+ \otimes E$. 

Then the non-Abelian Seiberg-Witten monopole equations in four 
dimensions \cite{LM2,LM3,HPP1,HPP2} are defined by
	\beqa
	F_{\mu \nu}^{a+} +\frac{i}{2}\ovM \sigma_{\mu \nu}T^a M &=&0 ,\nm \\
	i\gamma^{\mu}D_{\mu}M &=&0 ,\lab{mono}
	\eeqa
where 
	\beqa
	F_{\mu \nu}^{a+} &=& P_{\mu\nu\rho\sigma}^{+}F^{\rho\sigma a},\nm \\ 
	F_{\mu\nu}^{a} &=&\pa_{\mu} A_{\nu}^a -\pa_{\nu}A_{\mu}^a +f_{abc}
	A_{\mu}^b A_{\nu}^c .
	\eeqa
	
We take the classical action
	\beq
	S_c =\frac{1}{4}\int_X d^4 x \left[ 
	\left(G_{\mu \nu}^{a+} -F_{\mu \nu}^{a+} -\frac{i}{2}
	\ovM \sigma_{\mu \nu}T^a M\right)^2 +
	2|\nu -i\gamma^{\mu}D_{\mu}M|^2  
	\right] , 
	\lab{start}
	\eeq
where $G_{\mu \nu}^{a+}$ is a self-dual auxiliary field satisfying 
	\beq
	G_{\mu \nu}^{a+} = P_{\mu\nu\rho\sigma}^{+}G^{\rho \sigma a}.
	\eeq
The minimum of (\ref{start}) is then given by  
	\beqa
	G_{\mu\nu}^a -F_{\mu \nu}^{a+}-\frac{i}{2}\ovM \sigma_{\mu \nu}T^a 
	M &=&0 ,\nm \\
	\nu -i\gamma^{\mu}D_{\mu}M &=&0 .\lab{Langevin}
	\eeqa

The gauge symmetry of (\ref{start}) is 
	\beqa
	\delta A_{\mu}^a &=&\pa_{\mu} \theta^a +f_{abc}A_{\mu}^b \theta^c 
	+\epsilon_{\mu}^a , \nm \\
	\delta M&=&i\theta M +\varphi, \nm \\
	\delta G_{\mu\nu}^a &=&f_{abc}G_{\mu\nu}^b \theta^c +P_{\mu\nu\rho
	\sigma}^{+}\left[
	\pa^{[\rho}\epsilon^{\sigma ]a}+f_{abc}\epsilon^{b[\rho}  
	A^{\sigma ]c}+\frac{i}{2}(\overline{\varphi}
	\sigma^{\rho \sigma}T^a M+\ovM \sigma^{\rho\sigma}T^a 
	\varphi )\right], \nm\\
	\delta \nu&=& i\gamma^{\mu}D_{\mu}\varphi +\gamma^{\mu}
	\epsilon_{\mu} M +i\theta \nu .
	\lab{gaugealg}
	\eeqa

First stage reducibility of (\ref{gaugealg}) can be seen from the 
identification   
	\beq
	\theta^a =\Lambda^a,\ \epsilon_{\mu}^a =-\pa_{\mu}\Lambda^a -
	f_{abc}A_{\mu}^b \Lambda^c,\ 
	\varphi =-i\Lambda M.
	\eeq	
In fact, (\ref{gaugealg}) closes on-shell
	\beqa
	\delta A_{\mu}^a &=&0, \nm \\
	\delta M& =&0, \nm \\
	\delta G_{\mu\nu}^a &=& f_{abc}\Lambda^c\left[ G_{\mu\nu}^b 
	-P_{\mu\nu\rho\sigma}^{+}\left.
	\left(F^{\rho\sigma b}+\frac{i}{2}\ovM\sigma^{\rho\sigma}T^b 
	M\right)\right]\right|_{\mbox{\scriptsize on\mbox{-}shell}} =0, \nm \\
	\delta\nu &=& i\Lambda( \nu -i\gamma^{\mu}
	D_{\mu}M )|_{\mbox{\scriptsize on\mbox{-}shell}} =0 , 
	\lab{gaugealg2}
	\eeqa

In the $R$ coefficient and zero-eigenvector notation, we obtain 
	\beqa
	R_{\theta^b}^{A_{\mu}^a}
	&=&\pa_{\mu}\delta_{ab}+f_{acb}A_{\mu}^c ,\nm\\
	R_{\epsilon_{\nu}^b}^{A_{\mu}^a}&=&\delta_{ab}\delta_{\mu\nu} ,\nm\\
	R_{\theta^a}^M&=&iT^a M,\nm \\
	R_{\varphi}^M&=&1,\nm \\
	R_{\theta^b}^{G_{\mu\nu}^a}&=& f_{acb}G_{\mu\nu}^c ,\nm \\
	R_{\epsilon_{\alpha}^b}^{G_{\mu\nu}^a}&=&P_{\mu\nu\rho\sigma}^+ 
	(\pa^{[\rho}\delta^{\sigma ]\alpha}\delta^{ab}+
	f_{acd}\delta^{cb}\delta^{\alpha [\rho}A^{\sigma ]d} ),\nm \\
	R_{\varphi}^{G_{\mu\nu}^a}&=&\frac{i}{2}P_{\mu\nu\rho\sigma}^+ 
	\ovM \sigma^{\rho\sigma}T^a , \nm \\
	R_{\theta^a}^{\nu}&=&iT^a \nu, \nm \\
	R_{\varphi}^{\nu}&=&i\gamma^{\mu}D_{\mu},\nm \\
	R_{\epsilon_{\mu}^a}^{\nu}&=&\gamma^{\mu}T^a M  
	,\eeqa
and 
	\beq
	Z_{\Lambda^b}^{\theta^a}=\delta_{ab},\ Z_{\Lambda^b}^{\epsilon_{\mu}^a}
	=-\pa_{\mu}\delta_{ab} -f_{acb}A_{\mu}^c ,\ 
	Z_{\Lambda}^{\varphi}=-iM.
	\eeq

The assortment of ghost fields would be obvious again. 
Then we obtain the minimal solution to the master equation (see Appendix B)  
	\beq
	S(\Phi_{\mbox{\scriptsize min}},\Phi_{\mbox{\scriptsize min}}^* )=
	S_c +\int_X \mbox{tr}\ \Delta S d^4 x
	,\eeq
where 
	\beqa
	\Delta S&=&A_{\mu}^{*}(D^{\mu}c+\psi^{\mu})+
	M^* (icM+N)+\ovM^* (-ic\ovM +\ovN ) \nm\\
	& &+G_{\mu\nu}^{*}\widetilde{G}^{\mu\nu}
	-iN^{*} (\phi M+cN)+i\ovN^{*}(\phi\ovM+c\ovN)\nm \\
	& &+\nu^* (i\gamma^{\mu}D_{\mu}N +\gamma^{\mu}\psi_{\mu}M+
	ic\nu)+\overline{\nu}^* 
	\overline{(i\gamma^{\mu}D_{\mu}N+\gamma^{\mu}\psi_{\mu}M+ic\nu)}
	\nm\\
	& &+2i\nu^* \ovn^* \phi
	+\psi_{\mu}^{*}(-D^{\mu}\phi-i\{\psi^{\mu},c\}) 
	+c^{*}\left(\phi-\frac{i}{2}\{c,c\}\right)-i\phi^{*}[\phi,c] 
	\nm\\
	 & &-\frac{i}{2}\{G_{\mu\nu}^{*},G^{*\mu\nu}\}\phi 
	,\eeqa
where
	\beq
	\widetilde{G}_{\mu\nu}=i[c,G_{\mu\nu}]+P_{\mu\nu\rho\sigma}^+ 
	\left[D^{[\rho}\psi^{\sigma ]}
	+\frac{i}{2}(\overline{N}\sigma^{\rho \sigma}
	T_a T^a M+\ovM \sigma^{\rho\sigma}T_a T^a N)\right] 
	.\eeq
The equations 
	\beqa
	D_{[\mu}\psi_{\nu ]}+\frac{i}{2}(\overline{N}\sigma_{\mu\nu}
	T_a T^a M+\ovM \sigma_{\mu\nu}T_a T^a N) &=&0,\nm\\
	i\gamma^{\mu}D_{\mu}N+\gamma^{\mu}\psi_{\mu}M&=&0,
	\eeqa
can be seen as linearizations of the non-Abelian monopole equations and the 
number of linearly independent $\psi_{\mu}$ and $N$ gives the dimension of the 
moduli space ${\M}$ of 
solutions of (\ref{mono}). The dimension $d({\M})$ was found 
to be (for $SU(n)$ case) 
	\beq
	d({\M})=(4n-2)c_2 (E)-\frac{n^2 -1}{2}
	(\chi(X)+\sigma(X))-\frac{d_R}{4}\sigma(X)
	,\eeq
where $\chi$ and $\sigma$ are Euler number and signature of $X$, respectively, 
$c_2 (E)$ is the second Chern class of the 
representation bundle and $d_R$ is the dimension of the representation $R$ of 
the Lie algebra, but as we take $R$ to be fundamental representation, $d_R$ 
is identified with $n$ \cite{LM2,LM3,HPP1,HPP2}. 

We augment $\Phi_{\mbox{\scriptsize min}}$ by new fields $\chi_{\mu\nu}^a ,
d_{\mu\nu}^a ,\mu (\ovm),\zeta(\overline{\zeta}),\lambda,\rho,\eta,e$ 
and the corresponding anti-fields,
	\beq
	\begin{array}{rrrrrrrr}
	\chi_{\mu\nu}^a & \ d_{\mu\nu}^a&\ \mu&\ \zeta&
	\ \lambda&\ \rho&\ \eta&\ e \\
	-1^-&\ 0^+&\ -1^-&\ 0^+&\ -2^+&\ -1^-&\ -1^-&\ 0^+
	\end{array} 
	\eeq
with 
	\beq
	\begin{array}{cccc}
	\chi_{\mu\nu}^{*a}&\ \mu^*&\ \lambda^*&\ \rho^* \\
	0^+&0^+&1^-&0^+
	\end{array} 
	,\eeq
where $\chi_{\mu\nu}^a (\chi_{\mu\nu}^{*a})$ and $d_{\mu\nu}^a$ are 
self-dual. 
If we choose 
	\beqa
	G_{\mu\nu}&=&0,\nm\\
	\nu &=&0,\nm\\
	\pa^{\mu}A_{\mu} &=&0,\nm\\
	-D^{\mu}\psi_{\mu}+\frac{i}{2}(\ovN M-\ovM N)&=&0
	,\eeqa
we can obtain the non-Abelian four dimensional action compatible 
with that of Labastida and Mari\~{n}o \cite{LM1}. 
Then the gauge fermion is chosen to be   
	\beq
	\Psi =-\chi^{\mu\nu}G_{\mu\nu} -\ovm \nu -\mu \overline{\nu} 
	+\rho \pa^{\mu}A_{\mu} -
	\lambda\left[ -D^{\mu}\psi_{\mu}
	+\frac{i}{2}(\ovN M-\ovM N)\right].
	\eeq
After similar manipulations as have been done in the text, we find the 
quantum action to be 
	\beq
	S_q =S_c +\int_X \mbox{tr}\ \widetilde{\Delta}S d^4 x
	,\eeq
where
	\beqa
	\widetilde{\Delta}S&=& -\left[-D^{\mu}\psi_{\mu}+\frac{i}{2}(
	\ovN M -\ovM N)\right] \eta
	+\lambda[i\{\psi_{\mu},D^{\mu}c+\psi^{\mu}\}-D_{\mu}(D^{\mu}\phi +
	i\{\psi^{\mu},c\})]\nm\\
	& &+(\phi\ovM M -i\ovN N)\lambda \nm \\
	& &-\chi^{\mu\nu}P_{\mu\nu\rho\sigma}^+ 
	\left[D^{[\rho}\psi^{\sigma ]}
	+\frac{i}{2}(\ovN\sigma^{\rho \sigma}T_a T^a M +
	\ovM \sigma^{\rho\sigma}T_a T^a N)\right] \nm \\
	& &-\ovm (i\gamma^{\mu}D_{\mu}N+\gamma^{\mu}\psi_{\mu}M+ic\nu)
	+\overline{(i\gamma^{\mu}D_{\mu}N
	+\gamma^{\mu}\psi_{\mu}M+ic\nu)}\mu \nm \\
	& &+2i\phi \ovm\mu -\frac{i}{2}\{\chi_{\mu\nu},\chi^{\mu\nu}\}\phi +
	\rho(\pa^{\mu}D_{\mu}c+\pa^{\mu}\psi_{\mu})\nm \\
	& &-d^{\mu\nu}G_{\mu\nu}-\overline{\zeta}\nu -\ovn \zeta+e\pa^{\mu}
	A_{\mu} 
	.
	\eeqa
In this quantum action, setting 
	\beq
	M(\ovM)=N(\ovN)=\nu(\overline{\nu})=\mu (\ovm)=0,
	\eeq	
we can find that the resulting action coincides with 
that of Donaldson theory \cite{BBRT,LP} up to BRST-exact terms.

The BRST transformation will be 
	\beqa
	\delta_B A_{\mu}&=&-\epsilon(D_{\mu}c+\psi_{\mu}),\nm\\
	\delta_B M&=&-\epsilon(icM +N), \nm \\
	\delta_B G_{\mu\nu}&=&-\epsilon\left\{i[c,G_{\mu\nu}]
	-i[\chi_{\mu\nu},\phi]\right.\nm\\
	& &+\left.P_{\mu\nu\rho\sigma}^{+}\left[D^{[\rho}\psi^{\sigma ]}
	+\frac{i}{2}(\ovN \sigma^{\rho \sigma}T_a T^a M
	+\ovM \sigma^{\rho\sigma}T_a T^a N)\right]\right\}, \nm\\
	\delta_B \nu&=&-\epsilon(ic\nu+\gamma^{\mu}D_{\mu}N +\gamma^{\mu}
	\psi_{\mu}M -i\mu\phi ) ,\nm\\
	\delta_B c&=&\epsilon \left( \phi-\frac{i}{2}\{ c,c\}\right),\nm \\
	\delta_B \psi_{\mu}&=&-\epsilon\left(D_{\mu}\phi+i\{\psi_{\mu},c\}
	\right),\nm \\
	\delta_B \rho &=& \epsilon e,\nm \\
	\delta_B \lambda &=&-\epsilon \eta,\nm\\
	\delta_B \mu &=& \epsilon \zeta,\nm\\
	\delta_B N &=&-i\epsilon (\phi M+cN),\nm\\
	\delta_B \chi_{\mu\nu}&=&\epsilon d_{\mu\nu},\nm\\
	\delta_B \phi &=& i\epsilon [\phi,c] ,\nm\\
	\delta_B d_{\mu\nu}&=&
	\delta_B e =\delta_B \zeta =\delta_B \eta =0 
	.
	\eeqa

Note that we have enough fields to obtain 
precisely the Donaldson polynomials. They are the multiplet $(A_{\mu},
\psi_{\mu},\phi)$. However, again, we do not obtain any topological 
invariants associated with matter sector.

\begin{center}
\section*{Appendix B. Solution to master equation}
\end{center}

\renewcommand{\theequation}{B\arabic{equation}}\setcounter{equation}{0}
	
In this Appendix, the expansion coefficients in (\ref{coefficient}) 
for the non-Abelian case in four dimensions are determined. 
Since $B^{ji}_{\alpha}$ can be easily obtained, 
let us determine the structure functions $T_{\beta\gamma}^{\alpha}
C_{1}^{\gamma}C_{1}^{\beta}$. They are determined from the relation 
	\beq
	\frac{\pa_r R_{i}^{G_{\mu\nu}}C_{1}^i}{\pa \phi^k}
	R_{j}^{\phi^k}C_{1}^j +\left[R_{\theta}^{G_{\mu\nu}}
	T_{\beta\gamma}^{\theta}+R_{\epsilon_{\alpha}}^{G_{\mu\nu}}
	T_{\beta\gamma}^{\epsilon_{\alpha}}+
	R_{\varphi}^{G_{\mu\nu}} T_{\beta\gamma}^{\varphi}+
	R_{\overline{\varphi}}^{G_{\mu\nu}} T_{\beta\gamma}^{
	\overline{\varphi}}\right] C_{1}^{\gamma}C_{1}^{\beta}=0.
	\eeq
For example, $T_{\beta\gamma}^{\theta^b}$ are obtained from 
	\beq
	f_{acb}c^b f_{cef}G_{\mu\nu}^e c^f +
	f_{acb}G_{\mu\nu}^c T_{\beta\gamma}^{\theta^b}
	C_{1}^{\beta}C_{1}^{\gamma} =0.
	\eeq
The first term will be 
	\beqa
	f_{acb}f_{cef}G_{\mu\nu}^e c^b c^f &=&
	f_{adc}f_{deb}G_{\mu\nu}^e c^c c^b \nm\\
	&=&-(f_{abd}f_{cde}+f_{bcd}f_{ade})G_{\mu\nu}^e c^c c^b \nm\\
	&=&-f_{acd}f_{bde}G_{\mu\nu}^e c^b c^c -
	f_{bcd}f_{ade}G_{\mu\nu}^e c^c c^b \nm\\
	&=&f_{adc}f_{bde}G_{\mu\nu}^e c^b c^c 
	-f_{bed}f_{adc}G_{\mu\nu}^c c^e c^b \nm\\
	&=&-f_{adc}f_{bde}G_{\mu\nu}^e c^c c^b +
	f_{deb}f_{acb}G_{\mu\nu}^c c^e c^d ,
	\eeqa
thus we obtain 
	\beq
	f_{adc}f_{deb}G_{\mu\nu}^e c^c c^b =\frac{1}{2}f_{deb}f_{acb}
	G_{\mu\nu}^c c^e c^d .
	\eeq
Therefore,
	\beq
	T_{\beta\gamma}^{\theta^b} C_{1}^{\gamma}C_{1}^{\beta}=	
	\frac{1}{2}f_{bed}c^e c^d .
	\eeq
Since the other expansion coefficients can be obtained from similar 
calculations, 
we remain them as the reader's exercise. Final result will be 
	\beqa
	& &B_{\alpha}^{G_{\mu\nu}^a G_{\rho\sigma}^b}C_{2}^{\alpha}=
	\frac{f_{abc}}{2}\delta_{\mu\rho}\delta_{\nu\sigma}\phi^c ,\ 
	B_{\alpha}^{\nu\ovn}C_{2}^{\alpha}=-i\phi,\  
	B_{\alpha}^{\ovn\nu}C_{2}^{\alpha}=i\phi,\nm\\  
	& &T_{\beta\gamma}^{\varphi}C_{1}^{\gamma}C_{1}^{\beta}=-icN,\  
	T_{\beta\gamma}^{\overline{\varphi}}C_{1}^{\gamma}C_{1}^{\beta}=ic\ovN,
	\  T_{\beta\gamma}^{\epsilon_{\alpha}^a}C_{1}^{\gamma}
	C_{1}^{\beta}=f_{abc}\psi_{\alpha}^{b} c^c ,\nm\\
	& &A_{\beta\gamma}^{\Lambda^a}C_{1}^{\beta}
	C_{2}^{\gamma}=f_{abc}\phi^b c^c .
	\eeqa

\begin{center}
\section*{Appendix C. Convention for gamma matrix}
\end{center}

\renewcommand{\theequation}{C\arabic{equation}}\setcounter{equation}{0}

The convention for gamma matrix is as follows. Let $\sigma^i$ be 
Pauli matrices 
	\beq
	\sigma^1 =\left(\begin{array}{cc}
	0&1\\
	1&0\end{array}\right),\ 
	\sigma^2 =\left(\begin{array}{cc}
	0&-i\\
	i&0\end{array}\right),\ 
	\sigma^3 =\left(\begin{array}{cc}
	1&0\\
	0&-1\end{array}\right).
	\eeq
On $\RRRR$ we define four gamma matices 
	\beq
	\gamma^0 =\left(\begin{array}{cc}
	0&I\\
	I&0\end{array}\right)
	,\ 
	\gamma^j =\left(\begin{array}{cc}
	0&i\sigma^j\\
	-i\sigma^j &0\end{array}\right)
	\lab{gamma1}\eeq
and 
	\beqa
	\gamma^5 &=&\gamma^0 \gamma^1 \gamma^2 \gamma^3 \nm\\
		 &=&\left(\begin{array}{cc}
	I&0\\
	0&-I
	\end{array}\right),
	\eeqa 
where $I$ is a $2\times 2$ unit matrix. 

As can be easily seen from (\ref{gamma1}), they satisfy
	\beq
	\{\gamma^{\mu},\gamma^{\nu}\} =2\delta_{\mu\nu}
	.\eeq

It is often useful to define 
	\beq
	\sigma_{\mu\nu}=\frac{1}{2}[\gamma_{\mu},\gamma_{\nu}]
	.\eeq
Then 
	\beq
	\sigma_{ij}=i\epsilon_{ijk}
	\left(\begin{array}{cc}
	\sigma^k &0\\
	0&\sigma^k \end{array}\right),\ 
	\sigma_{k0}=i
	\left(\begin{array}{cc}
	\sigma^k &0\\
	0&-\sigma^k
	\end{array}\right).
	\eeq

On curved manifolds, we multiply viervein to these 
gamma matrices (except $\gamma^5$). On $Y\times [0,1]$, $\gamma^0$ 
is a constant matrix, while the others are not in general.

\begin{center}

\end{center}

\end{document}